\documentclass[pra,aps,twocolumn,showpacs]{revtex4}

\usepackage{bm}

\usepackage{graphicx}

\usepackage{color}

\begin{document}

\title{Experimental replication of single-qubit quantum phase gates}

\author{M. Mi\v{c}uda}
\affiliation{Department of Optics, Palack\'{y} University, 17. listopadu 1192/12, CZ-771 46 Olomouc, Czech Republic}

\author{R. St\'{a}rek}
\affiliation{Department of Optics, Palack\'{y} University, 17. listopadu 1192/12, CZ-771 46 Olomouc, Czech Republic}

\author{I. Straka}
\affiliation{Department of Optics, Palack\'{y} University, 17. listopadu 1192/12, CZ-771 46 Olomouc, Czech Republic}

\author{M. Mikov\'{a}}
\affiliation{Department of Optics, Palack\'{y} University, 17. listopadu 1192/12, CZ-771 46 Olomouc, Czech Republic}

\author{M. Sedl\'{a}k}
\affiliation{Department of Optics, Palack\'{y} University, 17. listopadu 1192/12, CZ-771 46 Olomouc, Czech Republic}

\author{M. Je\v{z}ek}
\affiliation{Department of Optics, Palack\'{y} University, 17. listopadu 1192/12, CZ-771 46 Olomouc, Czech Republic}

\author{J. Fiur\'{a}\v{s}ek}
\affiliation{Department of Optics, Palack\'{y} University, 17. listopadu 1192/12, CZ-771 46 Olomouc, Czech Republic}

\pacs{03.67.-a, 42.50.Ex, 03.65.Ud, 03.67.Hk}

\begin{abstract}
We experimentally demonstrate the underlying physical mechanism of the recently proposed protocol for superreplication of quantum phase gates 
[W. D{\"u}r, P. Sekatski, and M. Skotiniotis, Phys. Rev. Lett. \textbf{114}, 120503 (2015)],  which allows to produce up to $N^2$ high-fidelity replicas 
from $N$ input copies in the limit of large $N$.
 Our implementation of $1\rightarrow 2$ replication of the single-qubit phase gates is based on linear optics and qubits encoded into states of single photons. 
We employ the quantum Toffoli gate to imprint information about the structure of an input two-qubit state onto an auxiliary qubit, apply the replicated operation to the auxiliary qubit, 
and then disentangle the auxiliary qubit from the other qubits by a suitable quantum measurement. We characterize the replication protocol 
by full quantum process tomography and observe good agreement of the experimental results with theory.

\end{abstract}

\maketitle

\section{Introduction}

Laws of quantum mechanics impose fundamental limits on our ability to replicate quantum information \cite{Wootters82,Dieks82}. 
The quantum no-cloning theorem for instance ensures that quantum correlations cannot be exploited for superluminal signaling \cite{Dieks82}, and it guarantees the security of quantum key distribution \cite{Scarani09}. 
 These fundamental and practical impacts of the no-cloning theorem stimulated a wide range of  studies on approximate quantum cloning and
numerous  optimal quantum cloning machines for various sets of input states were designed and experimentally demonstrated \cite{Scarani05}.
 Recently, the concept of quantum cloning was extended from states to quantum operations \cite{Chiribella08,Bisio14}, which brings in new interesting aspects and features. 
 In particular, it was shown very recently that starting from $N$ copies of a single-qubit unitary operation $U$ 
one can deterministically generate up to $N^2$ copies of this operation with an exponentially small error \cite{Dur15,Chiribella15}.
As pointed out in Ref. \cite{Chiribella13}, the observed $N^2$ limit of the number of achievable almost perfect clones is deeply connected to the Heisenberg limit on quantum phase estimation.

In this paper we experimentally demonstrate the underlying physical mechanism of the superreplication of quantum gates, see Fig.~1. 
Specifically, we consider $1\rightarrow 2$ replication of single-qubit unitary phase gates \cite{Dur15}
\begin{equation}
U(\phi)=|0\rangle\langle 0| +e^{i\phi}|1\rangle\langle 1|, \qquad \phi\in [0,2\pi).
\label{Uphi}
\end{equation}
As shown in Fig.~1(b), we impose suitable quantum correlations between the two signal qubits and a third auxiliary qubit with the help of a three-qubit quantum Toffoli gate, 
we apply the operation $U(\phi)$ to the auxiliary qubit, and we finally erase the correlations between the auxiliary qubit and the signal qubits.
This latter part of the protocol can be in principle accomplished unitarily by a second application of the Toffoli gate, see Fig. 1(b), but we instead choose to erase
the correlations by means of a suitable measurement on the auxiliary qubit, see Fig.~1(c), which is much less demanding and makes the protocol experimentally feasible. 
Remarkably, this gate-replication scheme also simultaneously acts as a device which adds a control to an arbitrary unknown single-qubit phase gate 
and converts it to a two-qubit controlled phase gate \cite{Kitaev95,Lanyon09,Zhou11}.

\begin{figure}[!b!]
\includegraphics[width=\linewidth]{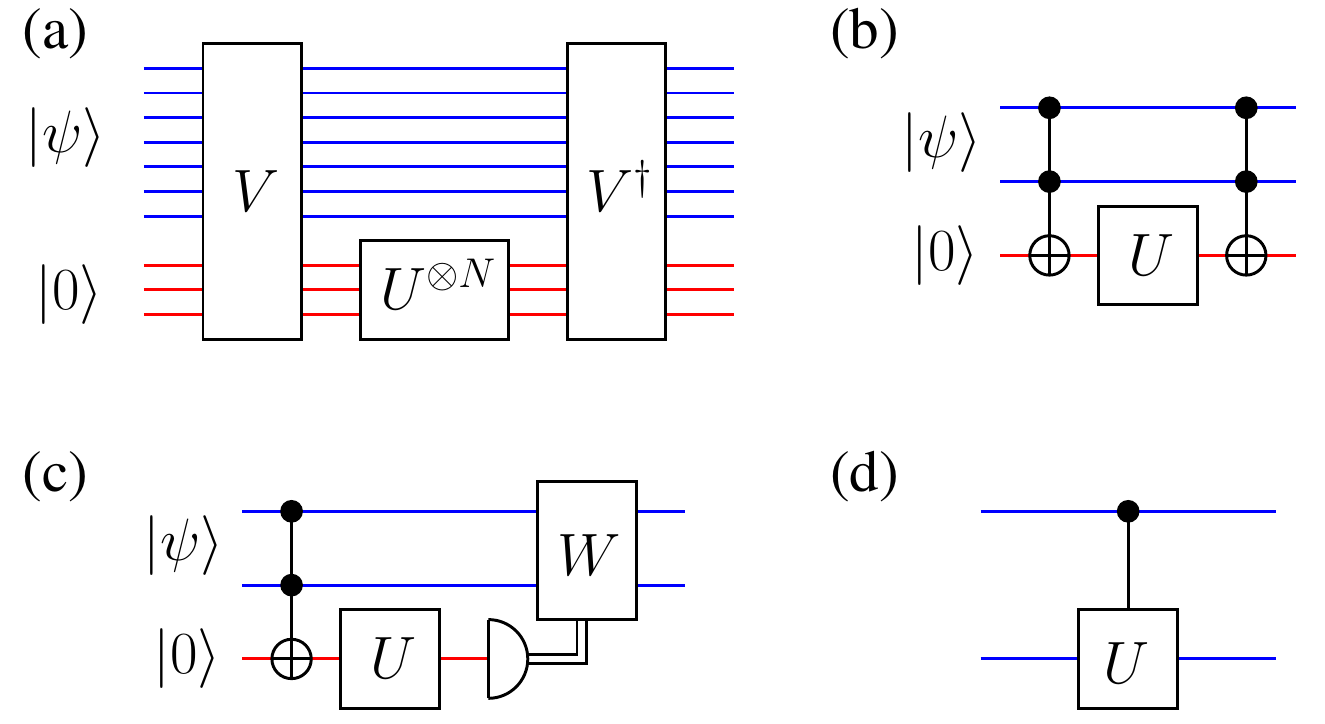}
\caption{(Color online) (a) Quantum circuit for superreplication of single-qubit quantum phase gates $U(\phi)$ \cite{Dur15}. (b)  $1\rightarrow 2$ quantum-gate replication circuit involving two quantum Toffoli gates and one ancilla qubit. 
(c) The same circuit as in panel (b), but  the second Toffoli gate is replaced with measurement on ancilla qubit and feed-forward. (d)
 The circuits in panels (b) and (c) convert a single-qubit phase gate $U(\phi)$ into a two-qubit controlled-phase gate. For details, see text.} 
\end{figure}

We experimentally implement the gate replication protocol using a linear optical setup, where qubits are encoded into states of single photons. We perform full quantum process
tomography of the replicated quantum gates and characterize the performance of the replication protocol by quantum gate fidelity. The linear optical quantum
Toffoli gate which we utilize \cite{Micuda13} is probabilistic and operates in the coincidence basis, hence requires postselection, 
similarly as other linear optical quantum gates \cite{Kok07}. 
Although the deterministic nature of the gate superreplication protocol is one of its important features,
our proof-of-principle probabilistic experiment still allows us to successfully demonstrate the underlying physical mechanism of quantum gate replication.

\section{Gate superreplication protocol}

Let us first briefly review the generic $N \rightarrow M$ superreplication protocol for phase gates $U(\phi)$ \cite{Dur15}.
We introduce the following notation for $M$-qubit computational basis states, 
$|\bm{m}\rangle=|m_1m_2\ldots m_M\rangle,$ where $\bm{m}\in[0,2^{M-1}]$ and $m_i\in \{0,1\}$ represent digits of binary representation of $\bm{m}$.
It holds that
\begin{equation}
U(\phi)^{\otimes M} |\bm{m}\rangle= e^{i|{\bm{m}}| \phi} |\bm{m}\rangle,
\end{equation}
where $|\bm{m}|=\sum_{i=1}^M m_i$ denotes the Hamming weight of an $M$-bit binary string $m_1m_2\ldots m_M$.
All basis states $|\bm{m}\rangle$ with the same Hamming weight $|\bm{m}|$ acquire the same phase shift $|\bm{m}|\phi$ and the number of such states $C_{|\bm{m}|}^M$
obeys binomial distribution, $C_{|\bm{m}|}^M= { M \choose |\bm{m}| }.$

This degeneracy is exploited in the  superreplication scheme illustrated in Fig. 1(a) \cite{Dur15}. The whole protocol can be divided into three steps. First, 
a unitary operation $V$ imprints information about the Hamming weight of the $M$-qubit signal state of system $A$ onto an auxiliary $N$-qubit system $B$, 
which is initially prepared in state $|\bm{0}\rangle_B$. In the joint computational basis $|\bm{m}\rangle_A|\bm{n}\rangle_B$ we explicitly have
\begin{equation}
\begin{array}{lcl}
V|\bm{m}\rangle_A |\bm{0}\rangle_B  = |\bm{m}\rangle_A |\bm{0}\rangle_B,&  \,\, & |\bm{m}|<m_{\mathrm{min}}, \\[2mm]
V|\bm{m}\rangle_A |\bm{0}\rangle_B  = |\bm{m}\rangle_A |\bm{k}(\bm{m})\rangle_B, &  \,\, & m_{\mathrm{min}}\leq |\bm{m}|< m_{\mathrm{max}}, \\[2mm]
V|\bm{m}\rangle_A |\bm{0}\rangle_B = |\bm{m}\rangle_A |\bm{2^N-1}\rangle_B,&  \,\, & |\bm{m}|\geq m_{\mathrm{max}}. 
\label{Voperation}
\end{array}
\end{equation}
Here each $|\bm{k}(\bm{m})\rangle$ is an $N$-qubit computational basis state chosen such that its Hamming weight satisfies $|\bm{k}|=|\bm{m}|-m_{\mathrm{min}}$, 
and the lower and upper thresholds on the Hamming weight are chosen symmetrically about the value $M/2$ for which $C_{|\bm{m}|}^M$ is maximized, 
$m_{\mathrm{min}}=\lceil\frac{M-N}{2}\rceil$, $m_{\mathrm{max}}=\lceil\frac{M+N}{2}\rceil$.

In the second step of the protocol, the $N$ replicated gates $U(\phi)^{\otimes N}$ are applied to the $N$ auxiliary qubits. Finally, in the third step, the system A is disentangled
from the auxiliary qubits by applying an inverse unitary operation $V^{-1}$. This sequence of operations results in the following unitary transformation 
on the $M$-qubit system A,
\begin{equation}
\begin{array}{lcl}
|\bm{m}\rangle   \rightarrow |\bm{m}\rangle&  \quad & |\bm{m}|<m_{\mathrm{min}}, \\[1mm]
|\bm{m}\rangle   \rightarrow e^{i(|\bm{m}|-m_{\mathrm{min}})\phi}|\bm{m}\rangle  &  \quad & m_{\mathrm{min}}\leq |\bm{m}|<m_{\mathrm{max}}, \\[1mm]
|\bm{m}\rangle  \rightarrow e^{iN\phi}|\bm{m}\rangle &  \quad & |\bm{m}|\geq m_{\mathrm{max}}. 
\end{array}
\label{superreplication}
\end{equation}
According to the Choi-Jamiolkowski isomorphism \cite{Choi75,Jamiolkowski72}, an $M$-qubit unitary operation $U$ can be represented by a pure maximally entangled state of $2M$ qubits,
\begin{equation}
|\Phi_U\rangle=I\otimes U |\Phi\rangle=\frac{1}{2^{M/2}}\sum_{\bm{m}=0}^{2^M-1} |\bm{m}\rangle \otimes U|\bm{m}\rangle, 
\label{PhiU}
\end{equation}
where $|\Phi\rangle=2^{-M/2}\sum_{\bm{m}=0}^{2^M-1} |\bm{m}\rangle|\bm{m}\rangle. $
Similarity of two unitary operations $U_1$ and $U_2$ can be quantified by the fidelity of the corresponding Choi-Jamiolkowski states,
$F_{U_1 U_2}=|\langle \Phi_{U_1}|\Phi_{U_2}\rangle|^2$, which can be expressed as  $F_{U_1U_2}=|\mathrm{Tr}[U_2^\dagger U_1]|^2/2^{2M}$.
It can be shown that in the limit of large $N$ the  fidelity of transformation (\ref{superreplication}) with the ideal operation $U(\phi)^{\otimes M}$ is exponentially close to one when 
 $M=N^{2-\alpha}$, $\alpha>0$.  Consequently, $N$ copies of $U(\phi)$ suffice to faithfully implement
up to $N^2$ copies of $U(\phi)$. An intuitive explanation for this result is that in the asymptotic limit the function $C_{|\bm{m}|}^M$ becomes Gaussian with a width that scales as $\sqrt{M}$.
If $M<N^2$, then the Hamming weights of an exponentially large majority of the basis states are located in the interval $m_{\mathrm{min}}\leq|\bm{m}|<m_{\mathrm{max}}$, 
and the superrepliction protocol induces correct relative phase shifts for all such states.

 Although experimental implementation of the superreplication protocol for large M is beyond the reach of current technology, the main mechanism of the superreplication 
 can be demonstrated already for $N=1$ and $M=2$. In this case we have $m_{\mathrm{min}}=1$ and $m_{\mathrm{max}}=2$, and it follows from Eq.~(\ref{Voperation}) 
 that $V$ is the three-qubit quantum Toffoli gate \cite{Nielsen00,Cory98,Monz09,Fedorov12,Reed12,Lanyon09,Micuda13}, which flips the state of the 
 auxiliary qubit if and only if both signal qubits are in state $|1\rangle$, see Fig. 1(b).
  Assuming a pure input state $|\psi_{\mathrm{in}}\rangle_A=c_{00}|00\rangle+c_{01}|01\rangle+c_{10}|10\rangle+c_{11}|11\rangle$ of the signal qubits
 and recalling that the auxiliary qubit is initially in state $|0\rangle$, the three-qubit state after the application of the Toffoli gate reads
 \begin{equation}
 |\Psi\rangle_{AB}=c_{00}|00\rangle|0\rangle+c_{01}|01\rangle|0\rangle+c_{10}|10\rangle|0\rangle+c_{11}|11\rangle|1\rangle.
 \end{equation}
 If we now apply the unitary operation $U(\phi)$ to the auxiliary qubit, and then disentangle this qubit from the rest by another application of the Toffoli gate,
 we obtain a pure output state of the two signal qubits,
 \begin{equation}
 |\psi_{\mathrm{out}}\rangle_A=c_{00}|00\rangle+c_{01}|01\rangle+c_{10}|10\rangle+e^{i\phi} c_{11}|11\rangle.
 \end{equation}
 The resulting two-qubit unitary operation thus reads
 \begin{equation}
 CU(\phi)=|00\rangle\langle 00|+|01\rangle\langle 01|+|10\rangle\langle 10|+e^{i\phi}|11\rangle\langle 11|.
 \label{CUgate}
 \end{equation}
The theoretical scheme in Fig. 1(b) can be further simplified by replacing the second Toffoli gate with measurement of the output auxiliary qubit in the superposition basis
 $|\pm\rangle=\frac{1}{\sqrt{2}}(|0\rangle\pm|1\rangle)$ followed by a feed-forward, see Fig.~1(c). In particular, a two-qubit controlled-Z gate should be applied to the signal qubits 
 if the auxiliary qubit is projected onto state $|-\rangle$ while no action is required when it is projected onto state $|+\rangle$.

 Fidelity of transformation (\ref{CUgate}) with the two replicas of the phase gate $U(\phi)\otimes U(\phi)$ reads
 \begin{equation}
 F_{UU}(\phi)= \frac{1}{8}(5+3\cos\phi).
 \label{FUUtheory}
 \end{equation}
 We note that the fidelity can be made phase independent and equal to a mean fidelity $\bar{F}_{UU}=\frac{5}{8}$ for all $\phi$ by twirling \cite{Bennett96},
 which would make the replication procedure phase-covariant. To apply the twirling, one has to generate a uniformly distributed random phase shift $\theta$  
 and apply a transformation $U(\theta)$ to the ancilla qubit, in addition to the replicated operation $U(\phi)$.  Subsequently, an inverse unitary operation 
 $U^\dagger(\theta)\otimes U^\dagger(\theta)$  is applied to the two output signal qubits.  Since $U(\phi)U(\theta)=U(\phi+\theta)$, it holds that
 for a fixed $\theta$ the fidelity of replication of $U(\phi)$ is given by Eq. (\ref{FUUtheory}), where $\phi$ is replaced with $\phi+\theta$. 
 After averaging over the random uniformly distributed $\theta$ we find that the fidelity becomes equal to the mean fidelity for any $\phi$.

 The mean fidelity $\bar{F}_{UU}$ exceeds the fidelity $\frac{1}{2}$ which is achievable by applying the transformation $U(\phi)$ 
 to one of the qubits while no operation is applied to the second qubit. Mean fidelity of $5/8$ can be also achieved by a measure-and-prepare scheme, 
 where $U(\phi)$ is applied to a fixed probe state $|+\rangle$, optimal phase estimation is performed on the output $U(\phi)|+\rangle$,
 and the estimated phase $\phi_{E}$ determines the operation $U(\phi_E)\otimes U(\phi_E)$ applied to the two signal qubits. 
 However, this latter procedure conceptually differs from the generic superreplication protocol and adds noise to the output state.

Remarkably, Eq. (\ref{CUgate}) shows that the quantum circuit depicted in Fig. 1(b) can be also interpreted as a scheme for deterministic conversion of an arbitrary 
 single-qubit unitary phase gate $U(\phi)$ into a two-qubit controlled unitary gate  $CU(\phi)$ \cite{Kitaev95,Lanyon09,Zhou11}.
While adding control to an arbitrary unknown single-qubit operation $U$ is known to be impossible \cite{Araujo14}, 
this task becomes feasible provided that one of the eigenvalues and eigenstates of the unitary operation is 
known \cite{Kitaev95}, which is precisely the case of the single-qubit phase gates (\ref{Uphi}) considered in the present work.

\begin{figure}[!t!]
\includegraphics[width=0.9\linewidth]{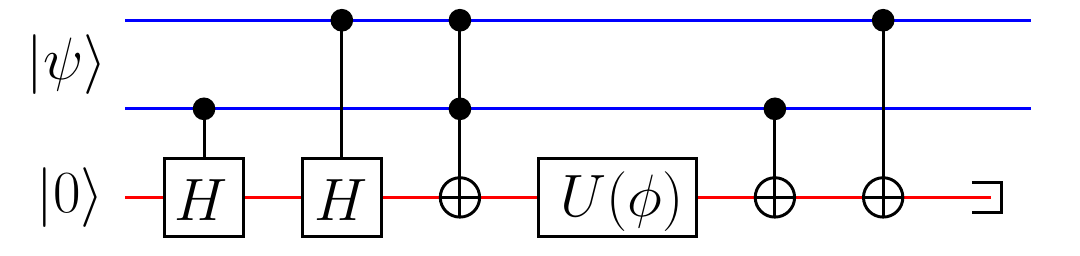}
\caption{(Color online) Quantum circuit for optimal $1\rightarrow 2$ replication of single-qubit unitary phase gates $U(\phi)$ \cite{Bisio14}.}
\end{figure}

An appealing feature of the superreplication protocol 
is that it is conceptually simple and universally applicable to any $N$ and $M$. 
Nevertheless, for specific $N$ and $M$ one can further optimize this scheme to maximize the replication fidelity. In particular, the optimal scheme for $1\rightarrow 2$ cloning of single-qubit phase gates
was recently derived by Bisio \emph{et al.} \cite{Bisio14} and is depicted in Fig.~2. Average fidelity of this optimal cloning procedure reads $F=(3+2\sqrt{2})/8\approx 0.729$. 
It is instructive to compare this scheme with the circuit in Fig.~1(b). Both circuits contain a Toffoli gate followed by application of $U(\phi)$ to ancilla qubit at their cores,
but the optimal cloning also requires two additional controlled-Hadamard (CH) gates on the input qubits and two controlled-NOT gates on the output qubits instead of a single Toffoli gate. 
The latter two CNOT gates can be replaced by a measurement of the  ancilla qubit in the superposition basis $|\pm\rangle$ followed by a suitable feed-forward on the signal qubits, 
similarly as in Fig.~1(c). However, the two CH gates are unavoidable, which makes the circuit in Fig.~2 much more difficult to implement than the circuit in Fig. 1(c).

\begin{figure}[!b!]
\includegraphics[width=\linewidth]{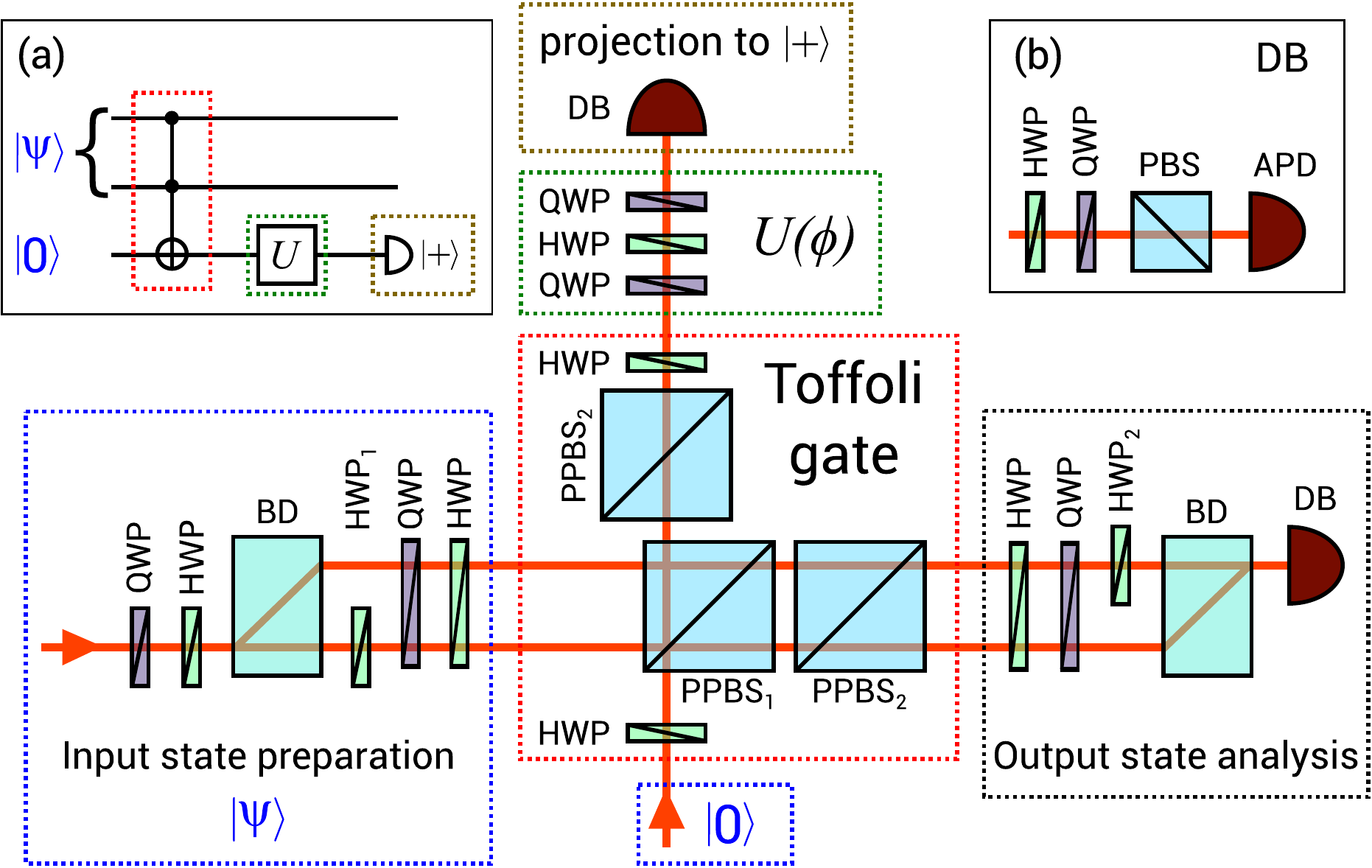}
\caption{(Color online) Experimental setup. HWP - half-wave plate, QWP - quarter-wave plate, 
PPBS - partially polarizing beam splitter with reflectances $R_V=2/3$ and $R_H=0$ for vertical and horizontal polarizations, respectively,
PBS - polarizing beam splitter, BD - calcite beam displacer, APD - single-photon detector.
 Inset (a) shows the implemented quantum circuit and inset (b) depicts the single-photon detection block DB.}
\end{figure}

 \begin{figure*}[!t!]
 \includegraphics[width=\linewidth]{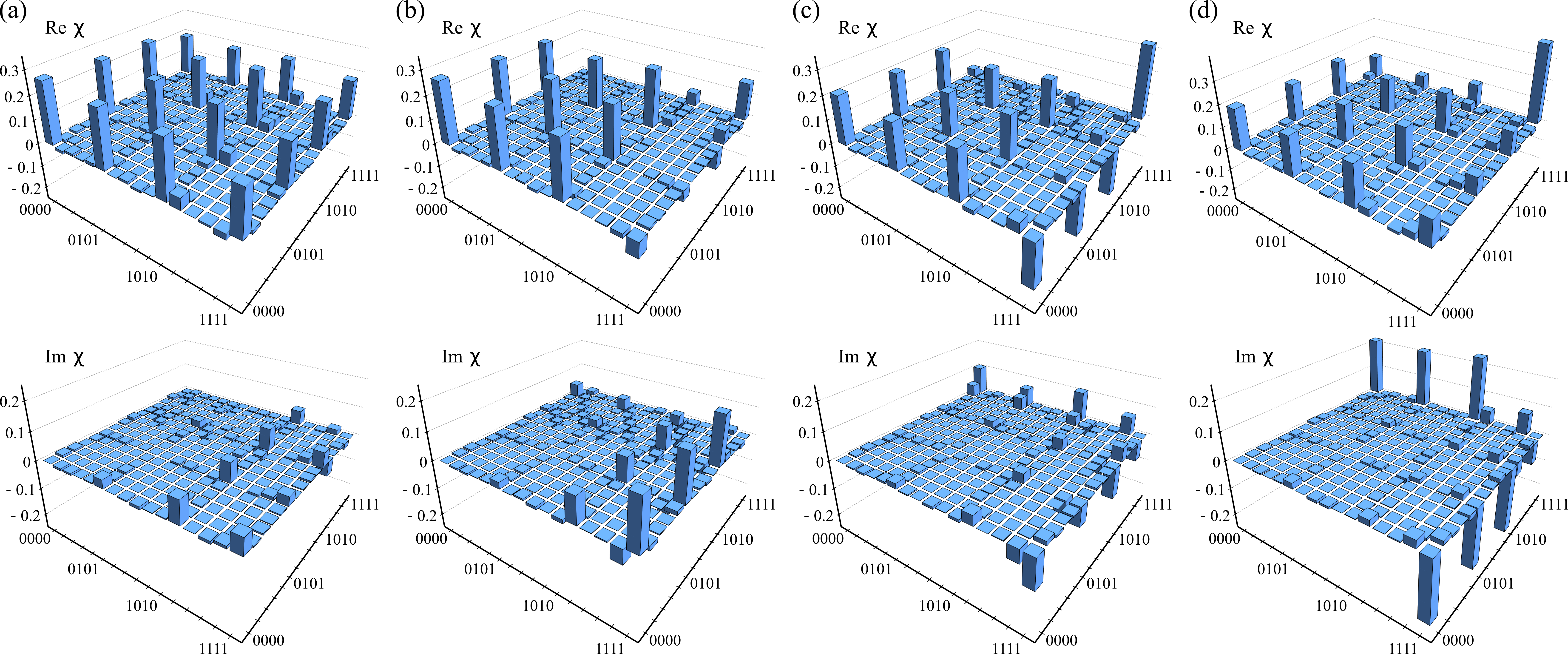}
 \caption{(Color online) Real and imaginary parts of the experimentally determined quantum process matrices $\chi$ representing the two-qubit operations $\mathcal{R}$ implemented on the two signal qubits 
 are plotted for four values of the phase shift:  $\phi=0$ (a),  $\phi=\pi/2$ (b),  $\phi=\pi$ (c),  and $\phi=3\pi/2$ (d). All process matrices are normalized such that $\mathrm{Tr}(\chi)=1$.}
 \end{figure*}

 \section{Experimental setup}
 
We experimentally implement the circuit in Fig.~1(c) which demonstrates the underlying physical mechanism of superreplication of quantum phase gates. 
The experimental setup is shown in Fig. 3 and its core is formed by our recently demonstrated linear optical quantum Toffoli gate \cite{Micuda13}. We utilize time-correlated photon pairs 
generated in the process of spontaneous parametric down-conversion
 in a nonlinear crystal pumped by a laser diode. The two signal qubits are encoded into the spatial and polarization degrees of freedom of the signal photon \cite{Fiorentino04,Gao10,Vitelli13,Rozema14}, 
 respectively, while the auxiliary qubit is represented by polarization state of the idler photon \cite{Micuda13,Rozema14}. 
 The spatial qubit is supported by an inherently stable Mach-Zehnder interferometer formed by two calcite beam-displacers BD which introduce a transversal 
 spatial offset between vertically and horizontally polarized beam. Arbitrary product input state of the three qubits can be prepared with the use of quarter- 
 and half-wave plates (QWP and HWP), and the output states can be measured in an arbitrary product basis with the help of a combination of wave-plates, polarizing beam splitters
 and single-photon detectors. The quantum Toffoli gate is implemented by a two-photon interference on a partially polarizing beam splitter PPBS$_1$ 
 that fully transmits horizontally polarized photons  and partially reflects vertically polarized photons  \cite{Okamoto05,Langford05,Kiesel05,Lemr11,Micuda13}. 
Similarly to other linear optical quantum gates \cite{Kok07}, this Toffoli gate operates in the coincidence basis \cite{Ralph02} 
and its success is indicated by simultaneous detection of a single photon at each output port. More details about the experimental setup can be found in Refs. \cite{Micuda13,Micuda15}.

The phase gate $U(\phi)$ on the polarization state of the idler photon is implemented using a sequence of suitably rotated quarter- and half-wave plates, see Fig.~3. 
At the output, the polarization of the idler photon is measured in the superposition basis $|\pm\rangle$, and we accept only those events where it 
is projected onto state $|+\rangle$. This ensures that we do not have to apply any feed-forward operation onto the signal qubits. 
While such conditioning reduces the success rate of the protocol by a factor of $2$, it does not represent a fundamental
modification, because the linear optical Toffoli gate utilized in our experiment is probabilistic in any case. 
Since both signal qubits are encoded into a state of a single photon, a real-time electrooptical feed-forward scheme \cite{Franson02,DeMartini02,Zeilinger07,Mikova12} 
could be in principle exploited to apply the controlled-Z  gate to signal qubits if the auxiliary idler qubit is projected onto state $|-\rangle$.

\section{Experimental results}

We have performed a full tomographic characterization of the experimentally implemented quantum gate replication protocol. 
The  measurements were performed for eight different values of the phase shift, $\phi=k\frac{\pi}{8}$, $k=0,1,\cdots,7$, and for each $\phi$
we have reconstructed the resulting quantum operation $\mathcal{R}$ on the two signal qubits.
We make use of the Choi-Jamiolkowski isomorphism \cite{Choi75,Jamiolkowski72} and represent  $\mathcal{R}$ 
by its corresponding quantum process matrix  $\chi =\mathcal{I}\otimes \mathcal{R} (|\Phi\rangle\langle \Phi|)$, where $|\Phi\rangle=\frac{1}{2}\sum_{j,k=0}^1|jk\rangle|jk\rangle$ 
denotes a maximally entangled state of four qubits, and $\mathcal{I}$ represents a two-qubit identity operation, c.f. also Eq. (\ref{PhiU}). 
  For each $\phi$, the process matrix  $\chi$ was reconstructed from the experimental data using a Maximum Likelihood estimation \cite{Hradil04}. 
  The experimentally determined matrices $\chi$ are plotted in Fig.~4 for four values of the phase shift $\phi$: $0$, $\frac{\pi}{2}$, $\pi$, and $\frac{3\pi}{2}$.
Additional results are provided in the Appendix which for comparison also contains
 plots of the corresponding theoretical matrices $\chi_{\mathrm{th}}$ of the ideal two-qubit unitary operations (\ref{CUgate}). 
 We quantify the performance of our experimental scheme by the quantum process fidelity $F_{CU}$ 
  of each implemented operation with the corresponding unitary operation $CU(\phi)$ that would be implemented by our setup under ideal experimental conditions. 
  Recall that fidelity of a general completely positive map $\chi$ with a unitary operation $U$ \cite{Schumacher96,Horodecki99} is defined as 
 $F=\langle \Phi_U|\chi | \Phi_U\rangle/\mathrm{Tr}[\chi]$.
 The fidelity $F_{CU}(\phi)$ calculated from the experimentally reconstructed process matrices $\chi$ is plotted in Fig. 5(a). 
 We find that $F_{CU}$ is in the range of $0.829 \leq F_{CU} \leq 0.897$
 and the mean fidelity reads $\bar{F}_{CU}=0.872$ which indicates high-quality performance of our setup for all $\phi$.

 \begin{figure}[!t!]
 \includegraphics[width=\linewidth]{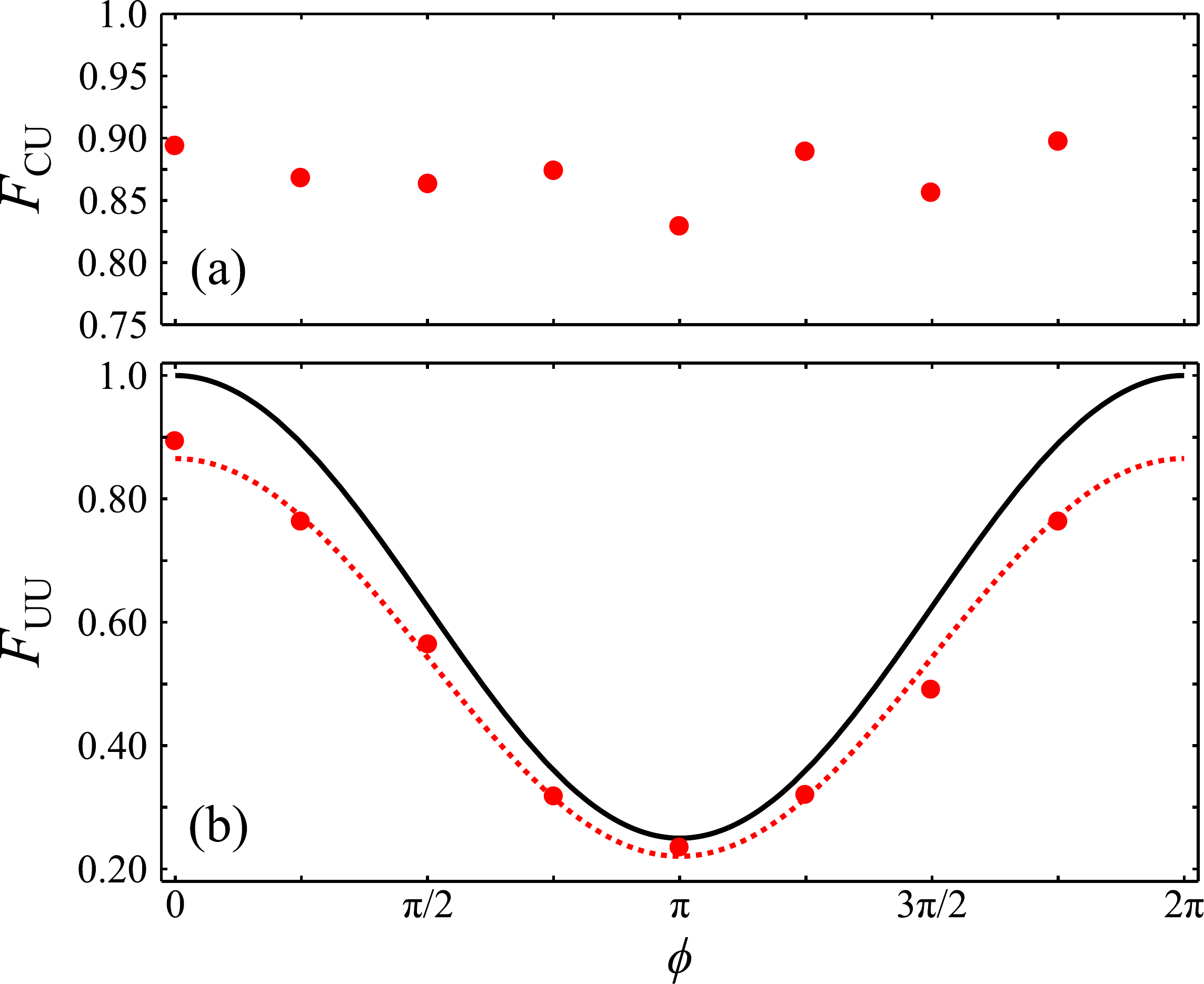}
 \caption{(Color online) Experimentally determined quantum gate fidelities $F_{CU}(\phi)$ (a) and $F_{UU}(\phi)$ (b) are plotted 
 for the eight values of $\phi$ that were probed experimentally. The solid line in panel (b) represents the ideal theoretical dependence (\ref{FUUtheory}) 
 and the dashed line is the least-square fit of the form $A+B\cos\phi$, with $A=0.543$ and  $B=0.322$. 
 Statistical error bars are smaller than the markers.}
 
 \end{figure}

The experimentally determined fidelities $F_{CU}$ are consistent with the fidelity of the Toffoli gate, $F_T=0.894$, which was determined in our earlier work \cite{Micuda15}. 
The main experimental limitations that reduce the fidelity of the protocol include imperfections of the central partially polarizing beam splitter PPBS$_1$, 
which exhibits $R_V=0.660$ and $R_H=0.017$  instead of $R_V=2/3$ and $R_H=0$, limited visibility of two-photon interference $\mathcal{V}=0.958$, 
and residual phase fluctuations in the Mach-Zehnder interferometer formed by the two calcite beam displacers \cite{Micuda15}. % cite also the weak CZ gate paper
Additional factors that may influence the experiment include imperfections of the wave plates and polarizing beam splitters 
that serve for state preparation and analysis. A more detailed discussion of these effects and their influence on performance 
of the experimental setup can be found in Ref. \cite{Micuda15}.

 We now turn our attention to characterization of gate replication. For this purpose, we calculate the fidelity $F_{U U} (\phi)$ 
 of the implemented operations with $U(\phi)\otimes U(\phi)$. The results are plotted in Fig. 5(b) and we can see that $F_{UU}$ exhibits the expected periodic dependence on $\phi$
as predicted by the theoretical formula (\ref{FUUtheory}). 
Due to various imperfections, the observed fidelities are smaller than the theoretical prediction,
and the dashed line in Fig. 5(b) shows the least-square fit of the form $A+B\cos\phi$ to the data, which well describes the observed dependence of $F_{UU}(\phi)$ on $\phi$. 
The mean fidelity reads $\bar{F}_{UU}=0.543$, which exceeds the fidelity $0.5$ achievable by a naive protocol where $U(\phi)$ is applied to one qubit while no operation is performed on the other qubit. 

Statistical uncertainties of fidelities were estimated assuming Poissonian statistics of the measured two-photon coincidence counts.
 Since the fidelities are estimated indirectly from the reconstructed quantum process matrices $\chi$, we have performed repeated
  Monte Carlo simulations of the experiment, and for each run of the simulation we have determined the quantum process matrix $\chi$ and the fidelities $F_{UU}$ and $F_{CU}$. 
  This yielded an ensemble of fidelities which was used to calculate the statistical errors. 
  The maximum statistical uncertainty of fidelity that we obtain reads 0.0013 (one standard deviation).

\vspace*{3mm}

\section{Summary}

In summary, we have successfully experimentally demonstrated the underlying physical mechanism of superreplication of quantum phase gates. 
Specifically, we have imprinted  information about the structure of the input state of signal qubits onto an auxiliary qubit via a suitable controlled unitary operation 
(here quantum Toffoli gate), applied the operation which should be replicated to the auxiliary qubit, and then disentangled the auxiliary qubit from the signal qubits by a suitable
quantum measurement.  
Intriguingly, this procedure converts the replicated single-qubit phase gate to a two-qubit controlled-phase gate so it adds 
a control to an arbitrary unknown phase gate $U(\phi)$.  

Our work paves the way towards experimental implementations of even more advanced schemes for quantum gate replication. The demonstrated approach based on the Toffoli gate 
can be extended to arbitrary number of signal qubits $M$ and copies of unitary phase operation $N$. For any $M$ and $N$, 
the transformation (\ref{Voperation}) could be implemented by a sequence of several 
$(M+N)$-qubit generalized Toffoli gates, where each generalized Tofolli gate would induce nontrivial bit flips on the subspace of auxiliary qubits for 
 one particular basis state of the signal qubits and would behave as an identity operation for all other basis states of the signal qubits. 
 However, for the relevant scenario $N \geq \sqrt{M}$ the number of gates would grow exponentially with the number of qubits.
Therefore, it is likely that the constituent generalized quantum Toffoli gates would have to exhibit fidelities exponentially close to $1$ 
to ensure the desired high-fidelity performance of the superreplication protocol.

\begin{acknowledgments}
This work was supported by the Czech Science Foundation (GA13-20319S). 

\end{acknowledgments}

\appendix*
\section*{Appendix: Plots of all reconstructed quantum process matrices}

\begin{figure*}
{\includegraphics[width=\linewidth]{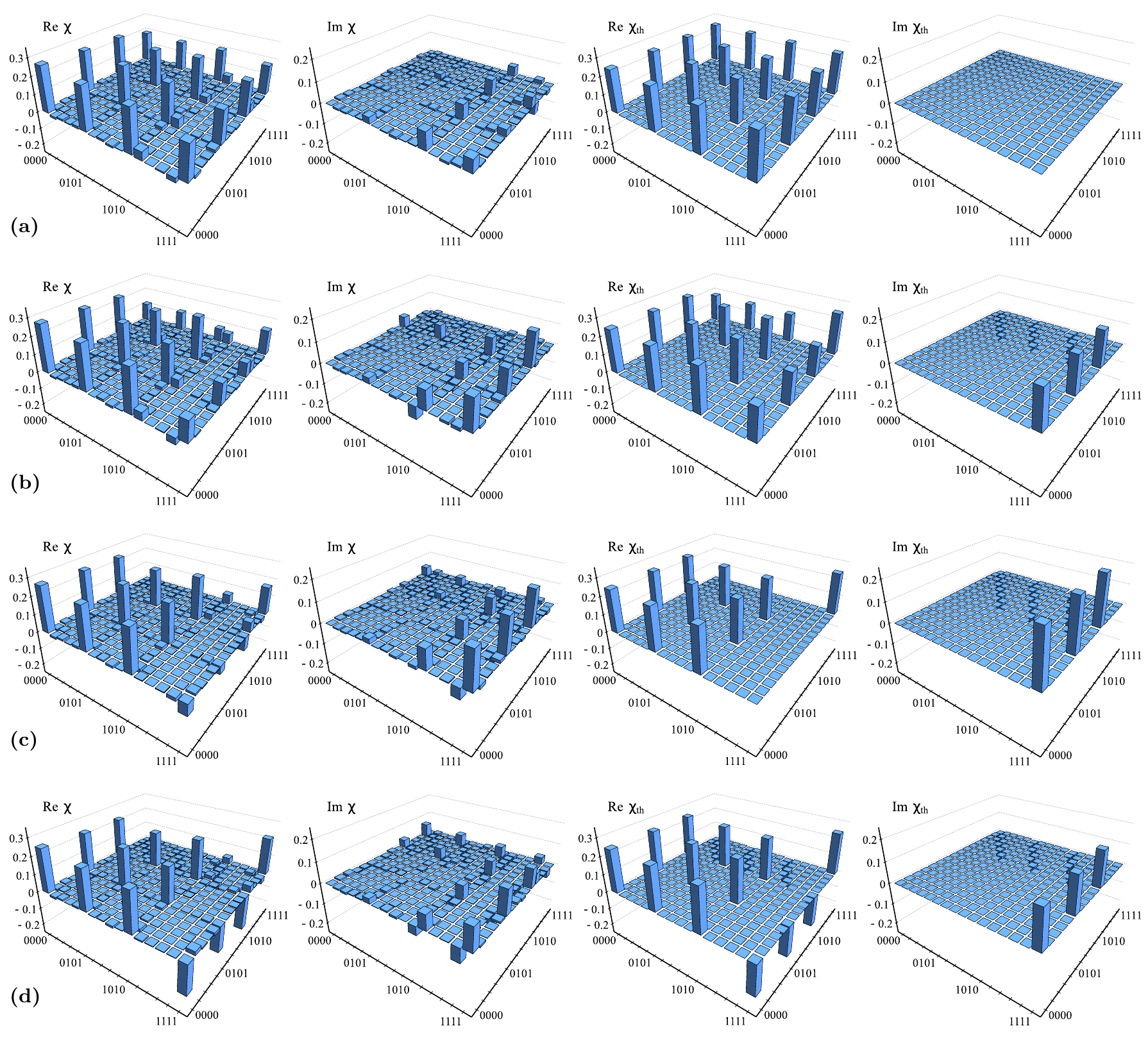}}
\caption{(Color online) Real and imaginary parts of the experimentally determined quantum process matrix $\chi$ representing the quantum operation implemented on the two signal qubits (first and second column), 
and real and imaginary parts of the quantum process matrix $\chi_{\mathrm{th}}$ of the corresponding ideal unitary operation $CU(\phi)$ (third and fourth column),  are plotted for four values of the phase shift: 
 $\phi=0$ (a),  $\phi=\pi/4$ (b),  $\phi=\pi/2$ (c),  and $\phi=3\pi/4$ (d). All process matrices are normalized such that $\mathrm{Tr}(\chi)=1$.}
 \label{afig1}
\end{figure*}

\begin{figure*}
{\includegraphics[width=\linewidth]{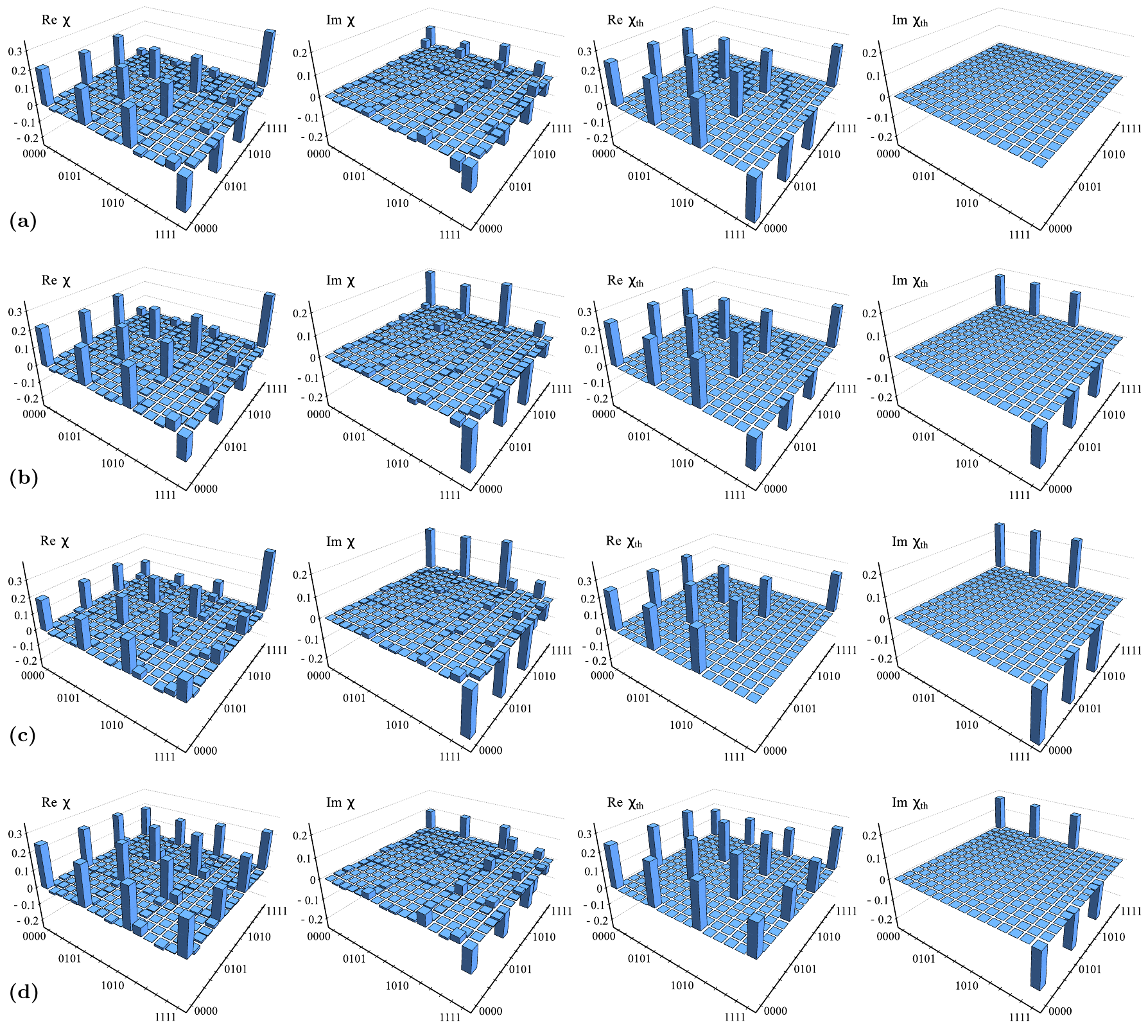}} 
\noindent\caption{(Color online) The same as Fig. \ref{afig1}, $\phi=\pi$ (a),  $\phi=5\pi/4$ (b),  $\phi=3\pi/2$ (c),  and $\phi=7\pi/4$ (d).}
 \label{afig2}
\end{figure*}

This Appendix contains two figures, Fig.~\ref{afig1} and Fig.~\ref{afig2}, that show all eight reconstructed quantum process matrices $\chi$ which characterize the quantum gates implemented on the two signal qubits for $\phi=k\frac{\pi}{8}$, $k=0,\ldots,7$.
For comparison, quantum process matrices $\chi_{\mathrm{th}}$ of the corresponding ideal unitary gates $CU(\phi)$ are also shown.

\begin{thebibliography}{99}

 
  \bibitem{Wootters82}
 W.-K. Wootters and W.-H. Zurek, Nature (London) \textbf{299}, 802 (1982).
 
 \bibitem{Dieks82}
D. Dieks, Phys. Lett. A \textbf{92}, 271 (1982).

\bibitem{Scarani09}
V. Scarani, H. Bechmann-Pasquinucci, N.J. Cerf, M. Du\v{s}ek, N. L\"{u}tkenhaus, and M. Peev, Rev. Mod. Phys. \textbf{81}, 1301 (2009).

  
 \bibitem{Scarani05}
 V. Scarani, S. Iblisdir, N. Gisin, and A. Ac\'{\i}n, Rev. Mod. Phys. \textbf{77}, 1225 (2005).
 
 
\bibitem{Chiribella08}
G. Chiribella, G.M. D'Ariano, and P. Perinotti, Phys. Rev. Lett. \textbf{101}, 180504 (2008).


\bibitem{Bisio14}
 A. Bisio, G.M. D'Ariano, P. Perinotti, and M. Sedl\'{a}k, Phys. Lett. A \textbf{378}, 1797 (2014).

 
 \bibitem{Dur15}
W. D{\"u}r, P. Sekatski, and M. Skotiniotis, Phys. Rev. Lett. \textbf{114}, 120503 (2015).
 
 \bibitem{Chiribella15}
G. Chiribella, Y. Yang, and C. Huang, Phys. Rev. Lett. \textbf{114}, 120504 (2015).



 \bibitem{Chiribella13}
 G. Chiribella, Y. Yang, and A. C.-C. Yao, Nat. Commun. \textbf{4}, 2915 (2013).
 

\bibitem{Kitaev95}
A. Y. Kitaev, quant-ph/9511026.

 
 \bibitem{Lanyon09}
B. P. Lanyon, M. Barbieri, M. P. Almeida, T. Jennewein, T. C. Ralph, K. J. Resch, G. J. Pryde, J. L. O'Brien, A. Gilchrist, and A. G. White, Nature Phys. \textbf{5}, 134 (2009).


 
\bibitem{Zhou11}
X.-Q. Zhou, T.-C. Ralph, P. Kalasuwan, M. Zhang, A. Peruzzo, B.P. Lanyon, and J.L. O'Brien, Nat. Commun. \textbf{2}, 413 (2011).

\bibitem{Micuda13}
M. Mi\v{c}uda, M. Sedl\'{a}k, I. Straka, M. Mikov\'{a}, M. Du\v{s}ek, M. Je\v{z}ek, and J. Fiur\'{a}\v{s}ek, 
Phys. Rev. Lett. \textbf{111}, 160407 (2013).


\bibitem{Kok07}
P. Kok, W.J. Munro, K. Nemoto, T.C. Ralph, J.P. Dowling, and G.J. Milburn, Rev. Mod. Phys. \textbf{79}, 135 (2007).


\bibitem{Jamiolkowski72}
A. Jamio\l{}kowski, Rep. Math. Phys. \textbf{3}, 275 (1972).

\bibitem{Choi75}
M.-D. Choi, 
Linear Algebra Appl. \textbf{10}, 285 (1975).



\bibitem{Nielsen00}
M. A. Nielsen and I. L. Chuang, \emph{Quantum Computation and Quantum Information} (Cambridge University Press, Cambridge, UK, 2000).


\bibitem{Cory98}
D. G. Cory, M. D. Price, W. Maas, E. Knill, R. Laflamme, W. H. Zurek, T. F. Havel, and S. S. Somaroo, 
Phys. Rev. Lett. \textbf{81}, 2152 (1998).

\bibitem{Monz09}
T. Monz, K. Kim, W. H\"{a}nsel, M. Riebe, A. S. Villar, P. Schindler, M. Chwalla, M. Hennrich, and R. Blatt, 
Phys. Rev. Lett. \textbf{102}, 040501 (2009).


\bibitem{Fedorov12}
A.~Fedorov, L.~Steffen, M.~Baur, M.~P.~da~Silva, and A.~Wallraff, 
Nature \textbf{481}, 170 (2012).

\bibitem{Reed12}
 M. D. Reed, L. DiCarlo, S. E. Nigg, L. Sun, L. Frunzio, S. M. Girvin, and R. J. Schoelkopf, 
 Nature \textbf{482}, 382 (2012).





\bibitem{Bennett96}
C. H. Bennett, G. Brassard, S. Popescu, B. Schumacher, J. A. Smolin, and W. K. Wootters, Phys. Rev. Lett. \textbf{76}, 722 (1996).


\bibitem{Araujo14}
M. Ara\'{u}jo, A. Feix, F. Costa, and \v{C}. Brukner, New J. Phys. \textbf{16}, 093026 (2014).



\bibitem{Fiorentino04}
M. Fiorentino and F.N.C. Wong, Phys. Rev. Lett. \textbf{93}, 070502 (2004).


\bibitem{Gao10}
W.-B. Gao, C.-Y. Lu, X.-C. Yao, P. Xu, O. G\"{u}hne, A. Goebel, Y.-A. Chen, C.-Z. Peng, Z.-B. Chen, and J.-W. Pan,
Nature Phys. \textbf{6}, 331  (2010)

\bibitem{Vitelli13}
C. Vitelli,	N. Spagnolo, L. Aparo, F. Sciarrino, E. Santamato, and L. Marrucci,	Nature Phot. \textbf{7}, 521 (2013).
        

\bibitem{Rozema14}
L.A. Rozema, D.H. Mahler, A. Hayat, P.S. Turner, and A.M. Steinberg, Phys. Rev. Lett. \textbf{113}, 160504 (2014).



\bibitem{Okamoto05}
R. Okamoto, H.F. Hofmann, S. Takeuchi, and K. Sasaki, Phys. Rev. Lett. \textbf{95}, 210506 (2005).


\bibitem{Langford05}
N. K. Langford, T.J. Weinhold, R. Prevedel, K. J. Resch, A. Gilchrist, J. L. O'Brien, G. J. Pryde, and A. G. White, 
 Phys. Rev. Lett. \textbf{95}, 210504 (2005).

\bibitem{Kiesel05}
N. Kiesel, C. Schmid, U. Weber, R. Ursin, and H. Weinfurter, Phys. Rev. Lett. \textbf{95}, 210505 (2005).


\bibitem{Lemr11}
K. Lemr, A. \v{C}ernoch, J. Soubusta, K. Kieling, J. Eisert, and M. Du\v{s}ek, 
Phys. Rev. Lett. \textbf{106}, 013602 (2011).




\bibitem{Ralph02}
T. C. Ralph, N. K. Langford, T. B. Bell, and A. G. White, 
Phys. Rev. A \textbf{65}, 062324 (2002).


\bibitem{Micuda15}
M. Mi\v{c}uda, M. Mikov\'{a}, I. Straka, M. Sedl\'{a}k, M. Du\v{s}ek, M. Je\v{z}ek, and J. Fiur\'{a}\v{s}ek, Phys. Rev. A \textbf{92}, 032312 (2015).



\bibitem{Franson02}
T. B. Pittman, B. C. Jacobs, and J. D. Franson, Phys. Rev. A  \textbf{66}, 052305 (2002).

\bibitem{DeMartini02}
S. Giacomini, F. Sciarrino, E. Lombardi, and F. De Martini Phys. Rev. A  \textbf{66}, 030302(R) (2002).

\bibitem{Zeilinger07}
R. Prevedel, P. Walther, F. Tiefenbacher, P. Böhi, R. Kaltenbaek, T. Jennewein, and A. Zeilinger,
Nature  \textbf{445}, 65-69 (2007).

 
 \bibitem{Mikova12}
M. Mikov\'{a}, H. Fikerov\'{a},  I. Straka, M. Mi\v{c}uda,  J. Fiur\'{a}\v{s}ek, M. Je\v{z}ek, and  M. Du\v{s}ek,
Phys. Rev. A \textbf{85}, 012305 (2012).


\bibitem{Hradil04}
Z. Hradil, J. \v{R}eh\'{a}\v{c}ek, J. Fiur\'{a}\v{s}ek, and M. Je\v{z}ek, 
\emph{Maximum-Likelihood Methods in Quantum Mechanics}, Lect. Notes Phys. \textbf{649}, 59 (2004).



 
\bibitem{Schumacher96}
B.W. Schumacher, Phys. Rev. A \textbf{54}, 2614 (1996).


\bibitem{Horodecki99}
M. Horodecki, P. Horodecki, and R. Horodecki, Phys. Rev. A \textbf{60}, 1888 (1999). 



\end{thebibliography}
 \end{document}